\documentclass[reprint,amsmath,amssymb,aps,prab,floatfix]{revtex4-2}
\usepackage[english] {babel}
\usepackage{amsfonts,hyperref,graphicx,soul,natbib,mathrsfs}
\usepackage[euler]{textgreek}
\usepackage{epstopdf}
\usepackage{xcolor}
\usepackage{soul}
\usepackage{physics}
\usepackage{graphics}

\usepackage{dcolumn}
\usepackage{bm}

\begin{document}

\title{Longitudinal current profile reconstruction from wakefield response in plasmas and structures}

\author{R. Roussel}%
\email{roussel@ucla.edu}
\affiliation{Department of Physics and Astronomy, University of California, Los Angeles, California 90095-1547, USA}%
\author{G. Andonian}%
\affiliation{Department of Physics and Astronomy, University of California, Los Angeles, California 90095-1547, USA}%
\author{S. S. Baturin}%
\affiliation{Department of Electrical Engineering and Department of Physics, Northern Illinois University, DeKalb, IL 60115, USA}%
\author{J. B. Rosenzweig}%
\affiliation{Department of Physics and Astronomy, University of California, Los Angeles, California 90095-1547, USA}%

\date{\today}

\begin{abstract}
Present-day and next-generation accelerators, particularly for applications in driving wakefield-based schemes, require longitudinal beam shaping and attendant longitudinal characterization  for experimental optimization.
Here we present a diagnostic method which reconstructs the longitudinal beam profile at the location of a wakefield-generating source.
The methods presented derive the longitudinal profile of a charged particle beam solely from measurement of the time-resolved centroid energy change due to wakefield effects.
The reconstruction technique is based on a deconvolution algorithm that is fully generalizable to any analytically or numerically calculable Green's function for the wakefield excitation mechanism. This method is shown to yield precise features in the longitudinal current distribution reconstruction.
We demonstrate the accuracy and efficacy of this technique using simulations and experimental examples, in both plasmas and dielectric structures, and compare
to the experimentally measured longitudinal beam parameters as available.
The limits of resolution and applicability to relevant scenarios are also examined.

\end{abstract}

\maketitle

\section{\label{sec:intro}Introduction}

Advanced acceleration techniques based on beam-driven wakefields have produced unprecedented results in terms of achievable gradients, exceeding 10's of GV/m in plasma \cite{Litos:2015} and the GV/m threshold in dielectrics \cite{OShea:2016}.
Going beyond proof-of-concept, one must now look to optimization of the acceleration scheme. In this regard, maximizing the efficiency of energy transfer to the wakefield requires breaking the symmetry in the beam distribution by precision manipulation of its longitudinal profile \cite{Bane:1985}.
The characterization of the longitudinal profile in real time is therefore critical for performance optimization in wakefield accelerators.

There are a myriad of techniques to measure the bunch profile of high brightness, picosecond-scale beams, including the use of deflecting cavities \cite{Maxson:2017}, electro-optic sampling \cite{Berden:2004}, and interferometry of emitted radiation \cite{Happek:1991, Andonian:2009, Andrews:2014}.
In this paper, we present a complimentary method to analyze the bunch profile from the beam-driven wakefields in a structure or plasma. This method can be employed in a wide range of conceivable wakefield interaction scenarios. 
Beyond acceleration, beam-driven wakefields are already used for phase space manipulations, such as beam compression \cite{Zhao:2018}, linear ramp shaping \cite{Andonian:2017:WFramp} or dechirping \cite{Antipov:2014dechirper,POSTECH:DC}. The information gleaned from diagnostics employing these techniques can yield detailed bunch profile information that is synergistic with the signal reconstruction method we present here.

\begin{figure}[b]
	\centering
	\includegraphics[width=1.0\linewidth]{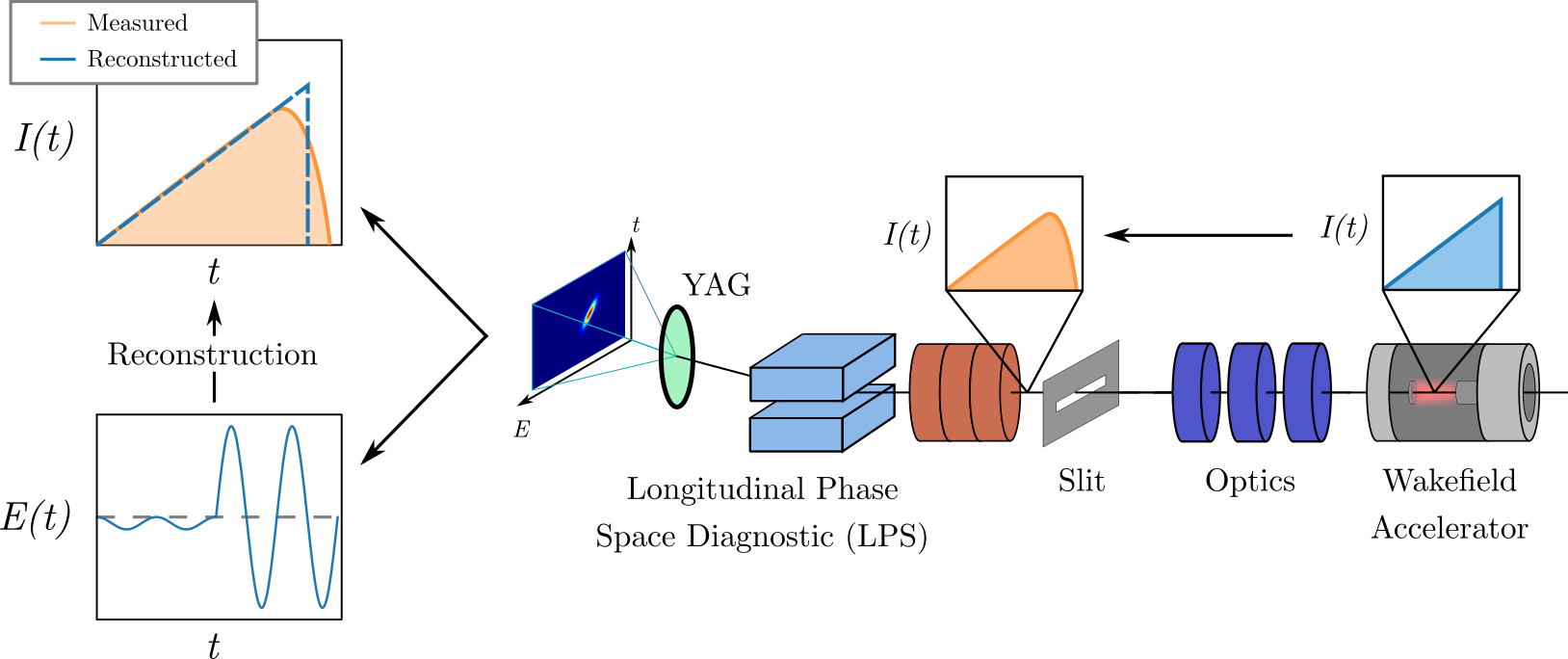}
	\caption{\label{fig:cartoon} Sketch of an example wakefield accelerator beamline (beam travels to the left) with a single stage wakefield accelerator and a single shot longitudinal phase space diagnostic. The reconstruction algorithm provides an accurate beam profile determination at the precise location of interaction, in contrast to the location of the diagnostic.}
\end{figure}

The bunch profile reconstruction technique we discuss in this work is crucially important for advanced acceleration applications because, unlike other methods, it provides the longitudinal beam profile at the immediate location of wakefield generation (Fig.\ref{fig:cartoon}).
Other diagnostic methods use beamline transport in an attempt to enhance the resolution obtained at the diagnostic location. 
However, it may not always be possible to carefully account for the effects of beamline elements on ps-scale features in the longitudinal profile during this transport.  
This creates challenges in obtaining accurate bunch information for wakefield optimization, because the beam longitudinal distribution evolves in a non-trivial way between the wakefield generating media and the available diagnostic.

For example, in the case of the emittance exchange \cite{EEX} process, which produces a strong correlation between the longitudinal profile and the vertical location of beam particles, the longitudinal bunch shape is modified by path length differences due to strong focusing quadrupole fields encountered as the beam particles traverse the diagnostic line \cite{Roussel:2019IPAC}. 
In this case, direct measurements of the longitudinal current profile far upstream or downstream of the wakefield accelerator are not reflective of the beam structure where wakefields are generated. 
While simulations can be used to provide an estimation of these effects, a direct measurement is often untenable, due to practical constraints (e.g. unwanted loss of beam particles and the information they carry) related to the wakefield device or diagnostic method used. The described reconstruction technique provides direct, single-shot, and real time longitudinal profile information with high precision, from the measured wakefield response. This method mitigates the above-described problems, providing a new path to robust and reliable beam profile measurements.

Furthermore, this method enhances the experimental accuracy obtained when employing a commonly encountered single shot longitudinal phase space diagnostic which consists of a transverse deflecting cavity and dipole spectrometer. 
In this diagnostic, a slit or pinhole is placed upstream of the deflecting cavity to reduce beam transverse emittances, and improve temporal and energy resolution.
However, in cases where the transverse and longitudinal phase spaces are strongly correlated, the improvement in resolution is a trade-off with overall profile accuracy because it only samples the near-axis beam current.

Finally, application of the reconstruction method yields profiles with high degree of precision, and even sub-resolution features beyond the diagnostic threshold, particularly when the wake functions of the media are well-understood.
In subsequent examples, it is shown that when these reconstructed profiles are employed interactively to obtain the experimental wakefield response, the predicted wakefield is accurately corroborated by the measured effects in the drive bunch and beyond, to the region left in its wake. 

\section{\label{sec:background}Background}
The longitudinal profile reconstruction technique discussed here is based on the signal deconvolution of time-resolved  energy change measurements due to the wakefield.
For a longitudinal wakefield, the drive beam initially experiences deceleration specific to the nature of the modes supported by the medium (dielectric, corrugated structure, or plasma).
This  manifests itself as a time-dependant change in beam energy, $\Delta U$, which is proportional to the induced wake potential, $W(s)$, which is defined as the convolution of the beam longitudinal distribution and the wake function,
\begin{align}
\label{eq:wake}
    W(s)=\int\limits_{0}^{s} \rho_z (\tilde s) G_z(s-\tilde s) d\tilde s, 
\end{align}
where $\rho_z(s)$ is the longitudinal beam distribution, and $G_z(s)$ is the Green's function representing the longitudinal wake function, or longitudinal wake potential of a point particle \cite{Chao}.
Eq.\eqref{eq:wake} is valid under the assumption of a pencil-like beam, where the transverse RMS beam size $\sigma_r$ is small compared to either the plasma wavelength or  vacuum channel size, in the case of a wakefield (\textit{e.g.} dielectric or metallic)  structure.
With known functions $W(s)$ and $G_z(s)$, a practical method of solving for $\rho_z(s)$ in this integral equation involves the application of Laplace transformations.

We define the forward and inverse Laplace transformations as \cite{shabat,silverman}

\begin{align}
\label{eq:lap}
\mathscr{L}[f(s)]=\int\limits_{0}^\infty f(s)e^{-ps}ds
\end{align}
and

\begin{align}
\label{eq:ilap}
\mathscr{L}^{-1}[F(p)] = \frac{1}{2 \pi i} \lim \limits_{R \to \infty} \int_{\kappa-iR}^{\kappa+iR} e^{s p} \ F(p) \ dp,
\end{align}
where $\kappa$ is a real number such that the line $p = \kappa$ in the complex plane avoids singularities of $F(p)$.
According to the convolution theorem for the Laplace transformation \cite{shabat}, the Laplace image of the convolution integral with the variable upper limit is equal to the product of the Laplace images of the the functions under the integral
\begin{align}
    \mathscr{L}\left[\rho_z*G_z\right] = \mathscr{L}\left[\rho_z(s)\right]
    \mathscr{L}\left[G_z(s)\right].
\end{align}
Here $*$ refers to a convolution, as shown in Eq.\eqref{eq:wake} $\rho_z*G_z\equiv \int\limits_{0}^{s} \rho_z (\tilde s) G_z(s-\tilde s) d\tilde s $.
After rearrangement, the charge density, $\rho_z(s)$, can be solved for by applying an inverse Laplace transformation on the quotient, as 
\begin{align}{\label{eq:general_algorithm}}
    \rho_z(s)= \mathscr{L}^{-1} \left[\frac{\mathscr{L}(W)}{\mathscr{L}(G_z)}\right].
\end{align}
In practice, the wake potential, proportional to the time-resolved energy change, is measurable with high precision, and the Green's function is analytically calculable for many types of media. 

In the generalized case when the Green's function $G(s)$ is given numerically, as are the measured values of $W(s)$, one can solve Eq.\eqref{eq:wake} approximately by replacing the convolution $\rho_z*G_z$ with the quadrature formula on a given mesh $s_n=nh$, with the mesh step $h$. The problem of Eq.\eqref{eq:wake} in this case is reduced to a linear system 
\begin{align}
\label{eq:numV}
    W(s_n)=\sum\limits_{j=m}^{n} A_m \rho_z(s_m) G_z(s_n-s_m)
\end{align}
that should be solved for $\rho_z(s_m)$.
Here, ${A_m}$ are coefficients of the chosen quadrature formula.
At this point, Eq.\eqref{eq:general_algorithm} and Eq.\eqref{eq:numV} are completely general, and the details of the beam reconstruction are dependent on the specific Green's function.

Hence, a longitudinal beam diagnostic method can be devised by selection of the medium with a well-known longitudinal wake response, and measurement of the beam energy change due to the self-induced wakefield. 
In many applications, such as wakefield acceleration or dechirping, the manipulating medium already provides the necessary change in beam energy to make the concept viable.
It is also notable that this method provides an independent measure of beam charge,  for proper normalization of  $G_z(s)$, with the integral $\int\rho_z(s)ds=Q$, where $Q$ is the total charge.  

In the following sections of this work, we explore the implications of this technique with specific application to reconstructing the bunch profile using different examples. 
First, we examine the simple case of a single mode wake function, which can be used in certain (linear and quasi-linear) cases of plasma wakefield acceleration \cite{Roussel:2020} and a limited scenario in dielectric wakefield acceleration. 
Then we extend the analysis to a more general approach that incorporates multiple modes, as well as arbitrary  Green's functions, which are essential for cases where the single-mode description cannot capture the complete wake response.
Experimentally-derived examples of reconstructed beam profiles are presented using time-resolved energy measurements from plasma and dielectric wakefield acceleration experiments at the Argonne Wakefield Accelerator \cite{Gao:2018}.
Further, we examine the fidelity of the diagnostic scheme under different situations and in the presence of measurement uncertainties. 

\section{\label{sec:theory}Single Mode Description} 

In the first case, we consider a single mode Green's function 
\begin{align}
\label{eq:SGF}
G_z(s)= G_0 \cos(k_0 s),
\end{align}
where $k_0$ is the wave-number associated with the fundamental mode.
We then apply Eq.\eqref{eq:general_algorithm} to derive the longitudinal profile $\rho_z(s)$. 
Consideration of this simple wake function is also relevant for cases such as a dielectric wakefield structure that can be optimized for a single-mode operation \cite{Bragg,Hoang:2018}, or quasi-linear plasma wakefields \cite{Roussel:2020}. 

The Laplace image of Eq.\eqref{eq:SGF} is
\begin{align}
    \mathscr{L}\left[G_0 \cos(k_0 s)\right]=G_0\frac{p}{p^2+k^2}.
\end{align}
The longitudinal profile is then  determinable from the inverse Laplace transformation as prescribed by Eq.\eqref{eq:general_algorithm}
\begin{align}{\label{eq:rho}}
    \rho_z(s) = \mathscr{L}^{-1}\left[
    \bar W(p)\frac{k_0^2+p^2}{G_0 p}
    \right]
\end{align}
where $\bar W(p)\equiv \mathscr{L}\left[W(s) \right]$.
By using the linearity of the Laplace transformation, we can expand Eq.\eqref{eq:rho} to obtain
\begin{align}{\label{eq:rhoi}}
    \rho_z(s) =\frac{1}{G_0}
    \mathscr{L}^{-1}\left[p
    \bar W(p) \right]+\frac{k_0^2}{G_0}\mathscr{L}^{-1}\left[ \frac{\bar W(p)}{p}  \right].
\end{align}
We notice that 
\begin{align}
\label{eq:L1}
    \mathscr{L}\left[W'(s) \right]=p\bar W(p)-W(0)
\end{align}
and 
\begin{align}
\label{eq:L2}
    \mathscr{L}\left[\int\limits_{0}^{s}W(\tilde s)d\tilde s \right]=\frac{
    \bar W(p)}{p}.
\end{align}
Here prime in $W'(s)$ denotes the total derivative.

Without loss of generality we chose the origin in $s$ such that $W(0)=0$. 
Finally, with Eq.\eqref{eq:L1} and Eq.\eqref{eq:L2}, Eq.\eqref{eq:rhoi} can be written as
\begin{align}{\label{eq:rhoii}}
    \rho_z(s) =\frac{W'(s)}{G_0}
    +\frac{k_0^2}{G_0}\int\limits_{0}^{s}W(\tilde s)d\tilde s.
\end{align}

As a check, in the simple case of constant $W(s)$, Eq.\eqref{eq:rhoii} represents a beam current density distribution that increases linearly, accompanied by a short current spike at the head of the beam. This is a familiar scenario found in theoretical treatments of beam-driven wakefield acceleration (e.g. \cite{Bane:1985,Baturin:2018,Lemery:2015,Baturin:TR,Li:2014,Baturin:BBU}). 

\subsection{Example: Plasma Wakefield Acceleration}
We apply the technique of Eq.~\ref{eq:rhoii} to the measurement of bunches in a plasma wakefield accelerator as part of a longitudinally-shaped-beam experiment described in Ref.~\cite{Roussel:2020}. This experiment was performed in a regime where the wakes inside the beam are described by linear plasma wakefield theory, where the beam density $n_b$ is much smaller than the plasma density $n_p$, the perturbation to the plasma density is given by  
\cite{Lu:2006,Rosenzweig:1988}.
\begin{equation}\label{eq:plasma_master}
    \frac{\partial^2 n_1(s)}{\partial^2 s^2}+k_p^2 n_1(s) = -k_p^2n_b(s),
\end{equation}
Furthermore, in this experiment, the beam is much narrower than the plasma skin-depth $k_p^{-1}=c/\omega_p$, and so the longitudinal wakefield is nearly constant across the radial extent of the beam. In the case of transition to the quasi-nonlinear regime, which is encountered in this experiment, the radial dependence of the longitudinal wake disappears completely.  Thus the representative Green's function for an axially symmetric beam is given by the familiar form
\begin{equation}
    G_z(s) = G_0 \cos(k_ps)
\end{equation}
where $k_p$ is the characteristic plasma wave-number $k_p$. From this, and an experimental measurement of the time dependent energy gain as a proxy for the average wakefield in the plasma, we reconstruct the longitudinal beam profile inside the plasma using Eq.\eqref{eq:rhoii}.

Some comments on the validity of this method as applied to the  ``quasi-nonlinear'' case are warranted.
The fully nonlinear plasma blowout is due to short ($k_p\sigma_z< 1$), high charge beams \cite{hogan_plasma_2010}. In contrast, in the experiments reported in Ref. \cite{Roussel:2020}, a long triangularly shaped beam is used to reach blowout conditions  through an extended build-up of beam charge density over several plasma wavelengths. In this case, the maximum beam charge density is larger than the unperturbed plasma density. 
With this strong perturbation electrons are  expelled from the beam channel, creating a rarefied ``bubble'' region around the beam where no plasma electrons are present.
However, the transverse motion of the plasma electrons is non-relativistic due to the adiabatic excitation method, and the maximum bubble radius $r_m$ is small compared to the plasma skin-depth. As a result, the longitudinal near-axis wakefield inside the drive bunch is dominated by fields generated by the population of plasma electrons reacting to the effectively linear perturbation to the plasma density outside the bubble \cite{lu_limits_2005}.  Thus the longitudinal wakefield response can be characterized as nearly linear,  and can be approximated by a single mode Green's function, of the same longitudinal dependence as the linear regime limit. 

\begin{figure}[t]
	\centering
	\includegraphics[width=1.0\linewidth]{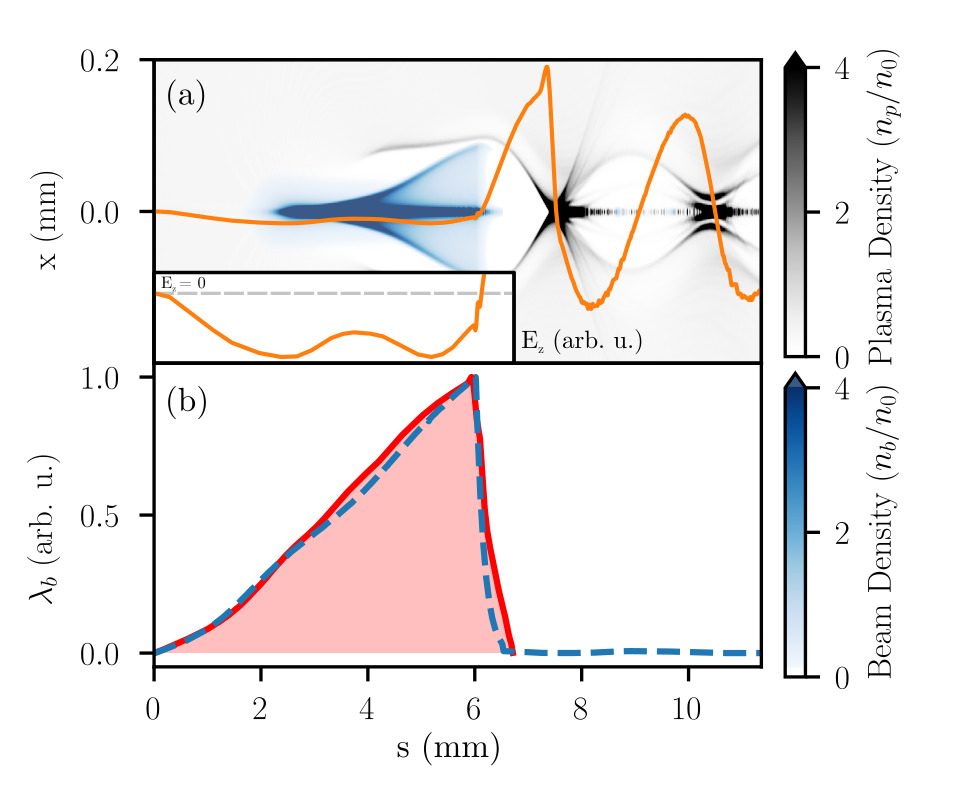}
	\caption{Warp simulation of a ramped beam after propagating 34 mm through a uniform plasma. (a) Longitudinal cross-section of the beam (blue) and plasma (grey) electron densities with the on-axis longitudinal electric field. Inset: Magnified view of the on-axis longitudinal wakefield. (b) Comparison between PIC simulated longitudinal current distribution (blue,dashed) and the single mode bunch reconstruction (red,filled) from the simulated wakefield $E_z$.}
\label{fig:plasma_pic}
\end{figure}

 \begin {table}[b]
\caption{Parameter for the WARP simulation and experiment from the Ref.\cite{Roussel:2020}}
\label{TB:pl}
\begin{ruledtabular}
\begin {tabular}{c c c c c c}
beam&charge&beam&beam spot&plasma&plasma \\
energy&&length&size&density&length \\
\colrule
40~MeV &1.8~nC&6~mm&200~$\mu$m&1.3$\times$10$^{14}$~cm$^{-3}$&8~cm 
\end {tabular}
\end{ruledtabular}
\end{table}  

This effect is demonstrated in Fig.\ref{fig:plasma_pic} through the use of the particle-in-cell code WARP \cite{Warp}. In the simulation, a linearly ramped beam with parameters matching those in Ref. \cite{Roussel:2020}, and displayed in Table \ref{TB:pl}, was injected into a uniform plasma with a density $n_0 = 1.3\times10^{14}\ \textrm{cm}^{-3}$ ($\lambda_p = 3$ mm). A longitudinal slice of the simulation is shown in Fig.\ref{fig:plasma_pic}(a) along with the on-axis longitudinal wakefield. Inside the drive beam, the wakefield is linear in nature as evidenced by its sinusoidal dependence on $s$ (see Fig.\ref{fig:plasma_pic}(a) inset). 
In Fig.~\ref{fig:plasma_pic}(b), the single mode reconstruction (Eq.\eqref{eq:rhoii}) technique is used to reconstruct the beam profile from the simulated wakefield. 
The reconstructed bunch profile shows excellent agreement with the simulated bunch profile, including reproduction of fine features such as the curvature at the beam head due to defocusing forces experienced by the beam before the bubble region fully develops.

It is important to stress that the reconstruction using a single mode Green's function is only accurate inside the drive beam for these experiments. In the beam region, the wakefield contribution is dominated by the quasi-linear plasma perturbation outside the bubble region. After the drive electron density suddenly drops  to zero (in a distance $<k_p^{-1}$), the bubble collapses with an attendant nonlinear response, and the wakefield is no longer approximated by a single, linear mode response.

\begin{figure}[t]
	\centering
	\includegraphics[width=0.85\linewidth]{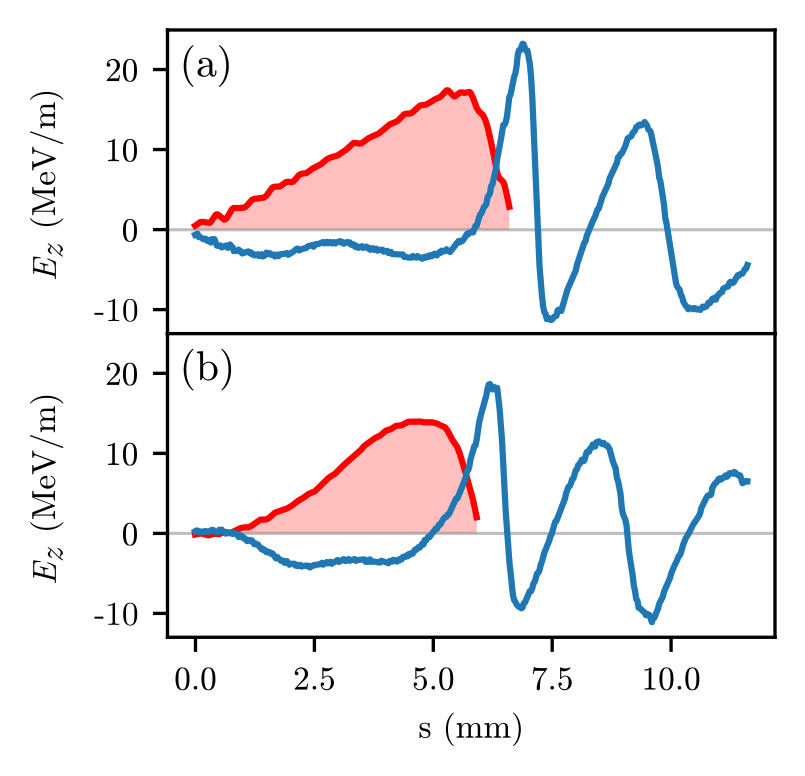}
	\caption{\label{fig:plasma_reconstruction} Wakefield measurement (blue) and beam current reconstruction (red, filled) from single shot longitudinal phase space measurements of the beam after a plasma accelerator. Measurements from two different beam shapes (a) linear ramp, (b) linear ramp with parabolic head are shown.}
\end{figure}

Demonstration of the reconstruction method for a quasi-nonlinear experimental case is plotted in Fig.\ref{fig:plasma_reconstruction}. 
In this experiment at the Argonne Wakefield Accelerator, an emittance exchange beamline \cite{EEX} was used to shape the temporal profile of the beam into two different ramped profiles, one without any leading perturbation and one with a parabolic head perturbation \cite{Roussel:2020}.
Here, the time-dependent energy change of the drive and witness bunches was measured using a single-shot longitudinal phase space diagnostic (Fig.~\ref{fig:cartoon}) \cite{Gao:PRAB}.
The time-dependent wakefield is extracted from the observed beam energy change due to the wakefield interaction averaged over the 8 cm long plasma column \cite{roussel_externally_2019}.
The plasma wave-number, $k_p$, is calculated by measuring the distance between adjacent minima in the wakefield in the drive region, which corresponds well to the plasma wavelength $2\pi/k_p$.
We apply our reconstruction method to the region before the first wakefield maximum, where the simulations indicate that we are operating in the quasi-nonlinear regime.
In the first case, Fig.\ref{fig:plasma_reconstruction}(a) shows a bunch with a near-linear ramp over roughly two plasma periods along with a short tail.
The development of a tail is consistent with beamline dynamics simulations as a symptom of the emittance exchange shaping process and strong focusing before the plasma stage  \cite{Roussel:2019IPAC}.
In Fig.\ref{fig:plasma_reconstruction}(b) we observe that the beam has a linear ramp with a parabolic head perturbation.
This perturbation has been shown analytically to flatten the decelerating wakefield inside the drive \cite{Lemery:2015} when the perturbation stretches over a plasma wavelength, which is consistent with our observations of the wakefield itself.

\section{Multi-mode Description}
The single-mode reconstruction technique works for a specific subset of problems, and is extremely useful for a rapid profile determination. 
However, for cases where the Green's function of the wakefield device is precisely known, either analytically or through simulation, a general approach can be used to expose finer features in the beam reconstruction. 
This includes both dielectric \cite{Andonian:2012} and corrugated metal structures and other media where the modal description is calculated numerically, such as the woodpile structure \cite{Hoang:2018} considered recently, where the wake function of a single particle is describable by a series of modes. 
In this section, we extend the applicability of the reconstruction algorithm beyond the single-mode description.

\subsection{Multi-mode Green's Function: Few-mode Case}
When extending the analysis to a multi-mode structure, the Green's function of Eq.\eqref{eq:SGF} can be replaced by a superposition of different mode contributions
\begin{align}
\label{eq:GFM}
G_z(s)=\sum\limits_{n=0}^{N} C_n \cos(k_n s).
\end{align}
First, we calculate the Laplace image of the Green's function using the linearity of the Laplace transformation
\begin{align}
    \bar G(p)=\mathscr{L}\left[ \sum\limits_{n=0}^{N} C_n \cos(k_n s) \right]= \sum\limits_{n=0}^{N} \frac{C_n p}{k_n^2+p^2}.
\end{align}

To find $\rho_z(s)$ according to the Eq.\eqref{eq:general_algorithm} and the definition of the inverse Laplace transformation Eq.\eqref{eq:ilap}  
the integral 
\begin{align}
\label{eq:rhoint}
\rho_z(s)=\frac{1}{2 \pi i} \int\limits_{\kappa-i\infty}^{\kappa+i\infty}  \frac{\bar W(p)e^{s p}\prod\limits_{n=0}^{N}(k_n^2+p^2)}{ p\sum\limits_{n=0}^{N}\left\{ C_n \prod\limits_{m\neq n}(k_m^2+p^2)\right\}} dp
\end{align}
needs to be evaluated.
An analysis of the integrand unveils an essential singularity at the points $p=\pm i\infty$ as well as poles that correspond to the zeros of the polynomial in the denominator:
\begin{align}
\label{eq:PEQ}
    p\sum\limits_{n=0}^{N}\left\{ C_n \prod\limits_{m\neq n}(k_m^2+p^2) \right\}=0.
\end{align}
The detailed evaluation of this integral is given in Appendix \ref{sec:appCI}.

Again taking the origin in $s$ such that $W(0)=0$ we arrive at the final expression for the charge density  
\begin{align}
\label{eq:MMF}
    \rho_z(s)&=\frac{W'(s)}{\sum\limits_{n=0}^{N} C_n}+\frac{\prod\limits_{n=0}^{N}k_n^2}{ \sum\limits_{n=0}^{N}\left\{ C_n \prod\limits_{m\neq n}k_m^2\right\}}\int\limits_{0}^{s} W(\tilde s) d\tilde s \nonumber \\ &+\int\limits_{0}^{s}K(s-\tilde{s})W(\tilde{s})d\tilde{s}.
\end{align}
with the kernel $K(s)$ given by
\begin{align}
\label{eq:MMfK}
K(s)=\sum\limits_{p_i} \Res[\frac{e^{s p}\prod\limits_{n=0}^{N}(k_n^2+p^2)}{ p\sum\limits_{n=0}^{N}\left\{ C_n \prod\limits_{m\neq n}(k_m^2+p^2)\right\}}].
\end{align}
Here $p_i$ are the roots of the Eq.\eqref{eq:PEQ} that are enclosed by the integration contour (see Appendix \ref{sec:appCI}) except for the root at $p=0$.
For the case of two and three modes, the residues in Eq.\eqref{eq:MMfK} can be evaluated analytically, as the equations for the poles Eq.\eqref{eq:PEQ} are quadratic and bi-quadratic equations respectively. It is still possible to evaluate the four-mode case but this involves Cardano formulas for the cubic equation. If the number of modes is greater then four, one should proceed with a numerical solution of Eq.\eqref{eq:PEQ}   
and substitution of the calculated zeros into the residue formula in Eq.\eqref{eq:MMfK}.  

\subsection{Arbitrary Green's function}
\label{Sec:ARG}
The evaluation of Eq.~\eqref{eq:wake} by a Laplace transformation is preferable as the integrals are calculated exactly, thus removing any error due to approximations. 
However, in some cases when the Green's function is computed using numerical simulation, it may be more convenient to consider a numerical reconstruction scheme. 
In this case the Green's function and the wake potential $W(s)$ are given on a mesh $s_n=nh$ where we assume that the mesh $s_n$ is uniform, with step size $h$.
Then we rewrite Eq.~\eqref{eq:wake} on this mesh as 
\begin{align}
\label{eq:wakeM}
    W(s_n)=\int\limits_{0}^{s_n} \rho_z (\tilde s) G_z(s_n-\tilde s) d\tilde s.
\end{align}    
If we assume that $h<<1$ then the integral on the right hand side can be approximated using the quadrature formula. If we chose the mesh for the integral $s_m=mh$ (the same as the initial mesh $s_n$) then we write the integral as
\begin{align}
\label{eq:QF}
    \int\limits_{0}^{s_n} \rho_z (\tilde s) G_z(s_n-\tilde s) d\tilde s\approx\sum\limits_{m=1}^{n} Q_m\rho_m G_{n,m}.
\end{align}
Here $Q_m$ are coefficients for the corresponding quadrature formula, $G_{n,m}=G_z(s_n-\tilde s_m)$ is the element of the square matrix $\mathrm G$, and $\rho_m=\rho_z (\tilde s_m)$ is the element of the unknown vector $\mathrm{\rho}$.  
Introducing the notation $W(s_n)\equiv W_n$ and using Eq.\eqref{eq:QF}, Eq.\eqref{eq:wakeM} becomes a square system of linear equations of the form
\begin{align}
\label{eq:ds}
    W_n=\sum\limits_{m=1}^{n} Q_m\rho_m G_{n,m}.
\end{align}
The most appropriate quadrature formula for the equidistant mesh and the case here is the Trapezoidal rule with coefficients $Q_m$ given by
\begin{align}
\label{eq:QCo}
    Q_m=\left\{
                \begin{array}{ll}
                  1/2~&m=1,n\\ 1~&m\neq 1,m\neq n.
                \end{array}
              \right.
\end{align}
The matrix $G_{n,m}$ has a triangular form, so all values of $\rho_m$ can be calculated using a simple recurrent formula that with Eq.\eqref{eq:QCo} for $k>2$ reads
\begin{align}
\label{eq:numr}
    \rho_k=2\frac{W_k-W_1-\frac{h}{2}\rho_1G_{k,1}-h\sum\limits_{l=2}^{k-1}\rho_lG_{k,l}}{hG_{k,k}}.
\end{align}
To initiate the calculations, we again assume $W(0)=0$ as well as $\rho_z(0)=0$. This allows us to  calculate $\rho_1$ as
\begin{align}
\label{eq:numr1}
    \rho_1=\frac{2W_1}{h G_{1,1}},
\end{align}
and $\rho_2$ as
\begin{align}
\label{eq:numr2}
    \rho_2=2\frac{W_2-W_1-\frac{h}{2}\rho_1G_{2,1}}{h G_{2,2}}.
\end{align}
If we assume that the total number of mesh points is $N$ and $\rho_0=\rho_z(0)=0$ then the total charge can be calculated using the same trapezoidal rule as follows 
\begin{align}
\label{eq:charge}
    Q=\frac{h\rho_N}{2}+h\sum\limits_{m=1}^{N-1}\rho_m.
\end{align}

\subsection{Example: Dielectric wakefield acceleration}

As a relevant example for the multi-mode case we consider the dielectric wakefield acceleration experiment described in Ref.\cite{Gao:2018}. In this experiment a triangular shaped bunch was generated using an emittance exchange beam line \cite{EEX} and sent into a slab dielectric structure. A single shot wakefield measurement system \cite{Gao:PRAB} was utilized to map out the wakefield behind this triangular shaped driver. Parameters of the dielectric slabs as well as external experimental conditions of Ref.\cite{Gao:2018} are summarized in Table \ref{TB:dl}.

 \begin {table}[b]
\caption{Experimental parameter of Ref.\cite{Gao:2018}}
\label{TB:dl}
\begin{ruledtabular}
\begin {tabular}{c c c c c c c}
beam&charge&vacuum&slab&width&length&$\varepsilon$ \\
energy&&gap&thickness&&& \\
\colrule
48~MeV &$\sim$2~nC&2.5~mm&150~$\mu$m&1.27~cm&15~cm&3.75 
\end {tabular}
\end{ruledtabular}
\end{table}

\begin{figure}[t]
	\centering
	\includegraphics[width=0.85\linewidth]{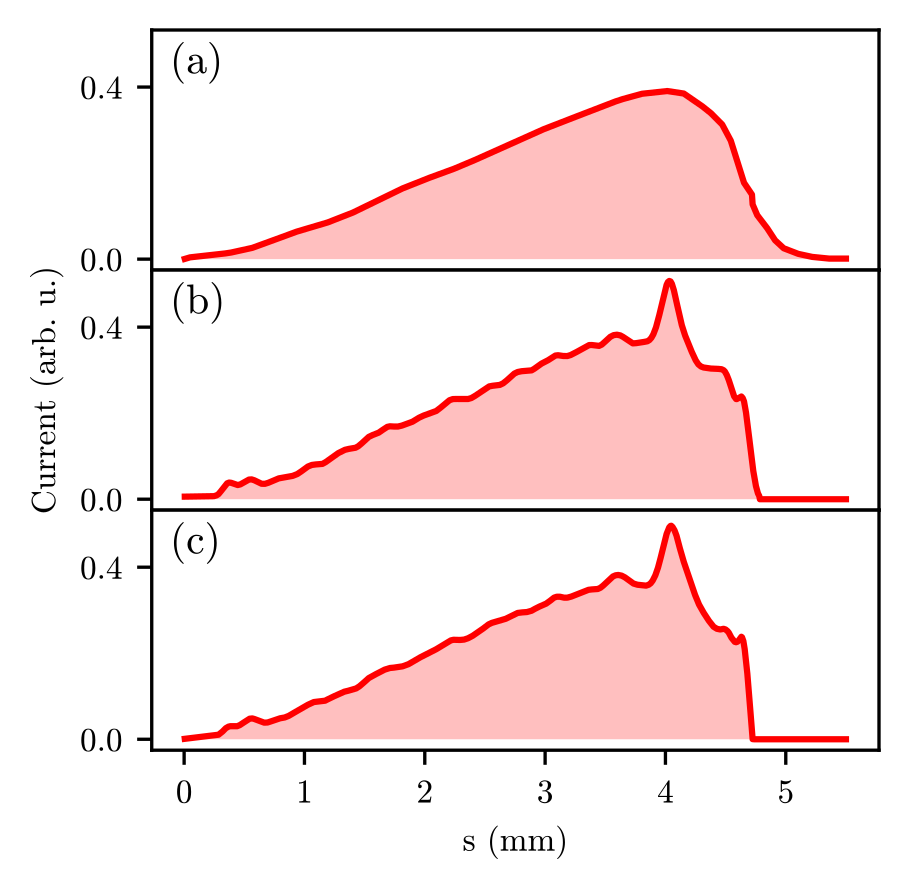}
	\caption{\label{fig:dielectric_reconstruction} (a) Normalized charge density from Ref.\cite{Gao:2018} measured with the LPS diagnostic. (b)  Normalized charge density at the wake field accelerating structure reconstructed using Eq.\eqref{eq:MMF} (c) Normalized charge density at the wake field accelerating structure reconstructed using Eq.\eqref{eq:numr}.}
\label{Fig:RDp}
\end{figure}
For a rectangular structure, the Green's function is represented by a double sum over $x$ and $y$ modes as \cite{Baturin:slab}
\begin{equation}
    G_z^{I}(s)=\sum\limits_{n_1}^{N_x}\sum\limits_{m=1}^{N_y} A_{n,m} \cos(k_{n,m}s)
\end{equation} 
(see Refs.\cite{jamie:rtg,piot:rtg,Baturin:slab}). In this specific case, to achieve high accuracy ($<0.1\%$) the Green's function has to include $N_x=80$ $x$-modes and $N_y=50$ $y$-modes, for a total of $4000$ modes. 
First, we apply Eq.\eqref{eq:MMF} to the experimentally measured wakefield inside the bunch.
We then apply a low-pass filter to reduce numerical noise that comes from the derivative term. 
The filter parameter was adjusted manually until the difference between convolution and measurement did not exceed 15$\%$, and most numerical artifacts were eliminated.
The bunch profile that resulted from the reconstruction performed in this manner is presented in Fig.\ref{Fig:RDp}(b) and compared to the charge density measured with the LPS diagnostic from Ref.\cite{Gao:2018}, Fig.\ref{Fig:RDp}(a). 
The reconstruction provides significant new insights into the fine features of the charge density profile as well as clarifying adjustments to the bunch length and the shape of the tail.

\begin{figure}[t]
	\centering
	\includegraphics[width=1.0\linewidth]{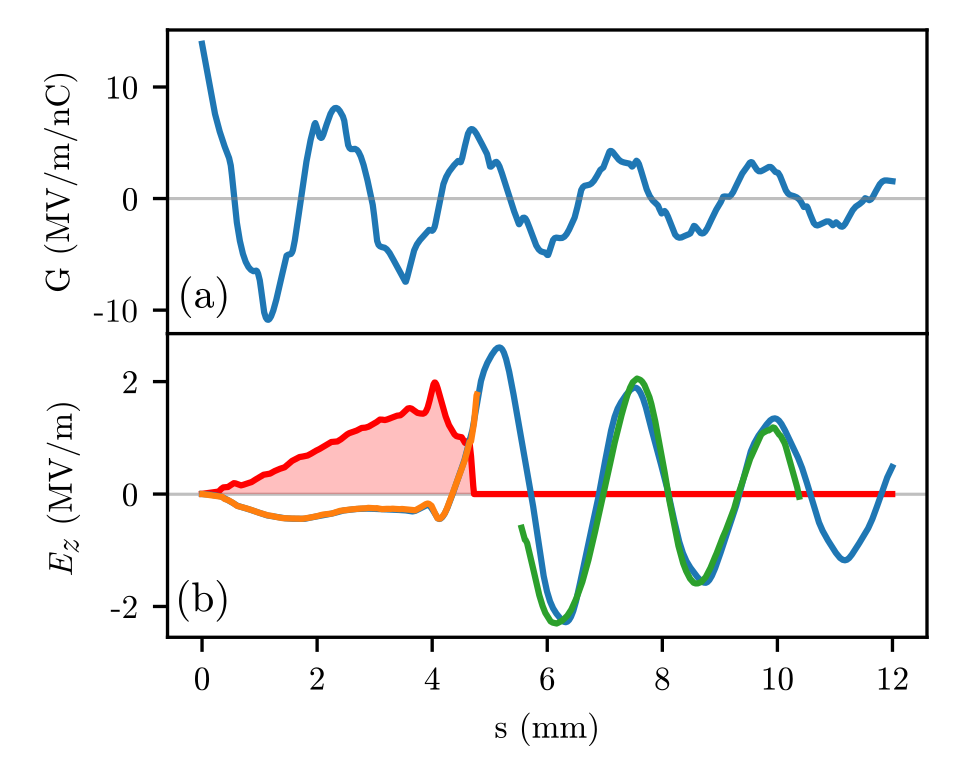}
	\caption{(a) Green's function for the structure that was used in Ref.\cite{Gao:2018} for measuring the wake field. (b) Reconstructed charge density (red, filled), measured wake field inside the driver (magenta), measured wake field behind the driver (green) and theoretical calculation of the full wake field based on a Green's function and reconstructed charge density (blue). }
\label{Fig:DLshow}
\end{figure}

Next, the Green's function was adjusted to account for the effects of the group velocity $v_g$ as $G_z(s)=G_z^{I}(s)\left[1-s/L(1-v_g)^{-1}\right]$. 
For the modes that give dominant contribution to the wakefield, $v_g\approx 0.8 c$ on average. 
The attenuated Green's function was evaluated numerically (Fig.\ref{Fig:DLshow}(a)) and then Eq.\eqref{eq:numr} with Eq.\eqref{eq:numr1} and Eq.\eqref{eq:numr2} were applied to restore the charge density profile. 
As before, numerical noise that comes from the discretization of the convolution integral was filtered out using low-pass filter. 
Filter parameters were optimized based on the same strategy as before, namely the reconstructed distribution was convoluted with the Green's function and compared with the initial measured wakefield inside the bunch; the filter parameter was then adjusted to reduce error between the convolution and the measurement. 
The largest errors occur at the bunch tail, however 15$\%$ accuracy is enough to remove the major part of the numerical noise while resolving the spike in the charge density closest to the tail. 
Further smoothing of the distribution and elimination of this spike results in a quite large 40-60 $\%$ difference between the measured value and the reverse convolution of the wake at these locations. 

We also use the reconstruction technique to evaluate the total charge by applying Eq.\eqref{eq:charge}. We calculate the total charge for the reconstructed profile to be $Q=1.8$ nC. This number is in agreement with the measured charge of $\sim 2$nC within the experimental errors, yielding an independent method to corroborate charge measurements.

It is important to note that profiles which are reconstructed using the two different approaches (Fig.\ref{fig:dielectric_reconstruction}(b) and Fig.\ref{fig:dielectric_reconstruction}(c)) match quite well. Slight differences at the tail of the distribution may be explained by slight differences in the Green's functions, as well as variation of the parameters of the low pass filter. 

In Fig.\ref{Fig:DLshow}(b) we compare the theoretical prediction of the wakefield generated by a reconstructed bunch with the measurements of the Ref.\cite{Gao:2018}. To produce the theoretical wakefield (blue line in Fig.\ref{Fig:DLshow}(b)) the Green's function for this structure Fig.\ref{Fig:DLshow}(a) is numerically convoluted with the reconstructed bunch profile (Fig.\ref{Fig:RDp}(a)).   
As expected, the measured wakefield inside the bunch (magenta line in Fig.\ref{Fig:DLshow}(b)) coincides with the theoretical prediction. This indicates that the reconstruction procedure has been successful and the bunch profile agrees with the measured wakefield. Furthermore, we observe that behind the bunch the calculated wakefield is in very close agreement with experimental observations (green line in Fig.\ref{Fig:DLshow}(b)) from Ref.\cite{Gao:2018}, providing further validation of this technique.

\section{Discussion}
It is apparent  from comparisons between the experimental and simulated wakefields that the reconstruction method accurately predicts the  drive beam profile present during the wakefield interaction. 
Several practical issues must carefully be taken into account when applying these reconstruction algorithms in order to reduce error. 
The accuracy of the reconstruction relies not only on the measurement accuracy of the wakefield, but also on the physical validity of the Green's function used to describe the wakefield response. For example, in the plasma wakefield case, the plasma wave-number was estimated by measuring the separation between adjacent wakefield minima in the drive beam (as well as from independent measurements of the plasma density).
Variations of the parameter $k_p$ in this case, changes the weighting between the derivative and integration terms in the Eq.\eqref{eq:rhoii}, leading to greater variations in the tail of the reconstructed beam, where the integration term is typically dominant. 

It is also important to consider that the single-mode reconstruction validity (Eq.~\eqref{eq:rhoii}) is extendable to certain multi-mode cases as well. By inspecting Eq.\eqref{eq:MMF} we conclude that if $\mathrm{max}|K(s)|$ given by Eq.\eqref{eq:MMfK} is much smaller than the multiplier in front of the integral in Eq.\eqref{eq:MMF} one may neglect the third convolution term and the reconstruction formula resembles Eq.\eqref{eq:rhoii}, except for multipliers in front of the integral and derivative. Interestingly these factors could be found by a simple optimization procedure with an \textit{ansatz} $\rho_z(s)=A\left(W^\prime(s)+B\int\limits_0^s W(\tilde s)d\tilde s\right)$, which is convoluted with the known Green’s function. Then, the constants $A$ and $B$ are varied to minimize $\mathrm{max} \left|W(s)-\int\limits_0^s\rho_z(\tilde{s})G_z(s-\tilde s) d\tilde s\right|$.

The smoothing strategy for the low-pass filter that was applied in the example cases relies on a similar optimization method.
The back-convolution of the filtered result is compared to the measured wakefield and the maximum modulus of the difference is kept below a certain threshold.
This strategy shows the trade-off between smoothness of the reconstructed profile and accuracy of the reconstruction procedure.

The universal solution introduced in Section \ref{Sec:ARG} has a built in error that is connected with the discretization Eq.~\eqref{eq:numV} of the convolution integral Eq.~\eqref{eq:wake}. 
Oversampling of the measured data for the $W(s)$ and subsequent filtering solves this problem, however analytical evaluation of Eq.\eqref{eq:general_algorithm} is preferable, as it does not produce additional errors and requires less aggressive filtering. 
This in turn enables finer resolution in certain scenarios.

Finally, if the transverse size of the beam is significant (close to the size of the vacuum gap of the structure or plasma skin-depth) then additional complicating considerations regarding transverse form-factor must be incorporated into the reconstruction procedure. 
This is critical for some linear PWFA experiments (as might be encountered at AWAKE \cite{Adli:2018}, for example). It is also  especially important in the multi-mode case as modal amplitudes have to be modified to account for the transverse beam shape and size. 

\section{Summary}

In this work, we have suggested an innovative approach to determining the longitudinal charge distribution, based on deconvolution of measured longitudinal wakefield effects in a beam.
The algorithm is generalizable to a wide variety of experimental scenarios and as such, is applicable to a variety of wakefield-based media (plasma, dielectric, corrugated, photonic and Bragg structures etc.). It permits, in these applications, reconstruction  of the bunch profile with high-precision, reproducing fine features that may not be observable with other beam measurement methods. 
This technique has been successfully applied to reconstruct longitudinal charge distribution in the plasma wakefield acceleration experiment described in Ref.\cite{Roussel:2020}, and in the dielectric wakefield acceleration experiment described in Ref.\cite{Gao:2018}.  
It is important to emphasize that the methods presented, whether they are single- or multi-mode, yield excellent results for the asymmetric beam shapes of highest interest. 
For rapid diagnosis of such cases, particularly in the PWFA, the single-mode treatment works very well.  
Enhanced beam features with finer resolution may require the multi-mode approach, while the universal solution proposed here has some inherent numerical fluctuations that must be filtered to obtain robust results. 

The  technique developed here may be incorporated into the diagnostic suite employed at any facility that has longitudinal phase space measurement capability, such as at upcoming experimental efforts at SLAC FACET-II \cite{FACET}, where detailed bunch profile information at the wakefield interaction location is needed for effective operation. 
Further, the method is single-shot and may be utilized in real-time, leading to ready-made applications in optimization of the current profile using novel machine learning feedback loops. These mechanisms can be further augmented with other diagnostics to provide accurate predictions for optimizing bunch properties for beam-driven wakefield acceleration. 

\begin{acknowledgments}
This work is supported by the Department of Energy, Office of High Energy Physics, under Grants No. DE-SC0017648 and No. DE-SC0009914. 
\end{acknowledgments}

\appendix

\section{Evaluation of the integral in \texorpdfstring{Eq.\eqref{eq:rhoint}}{Eq.(18)}}\label{sec:appCI}
In this appendix we present a detailed evaluation of the inverse Laplace transformation for the multi-mode case. For convenience we reproduce the inverse Laplace integral Eq.\eqref{eq:rhoint}:
\begin{align}
\label{eq:rhointA}
\rho_z(s)=\frac{1}{2 \pi i} \int\limits_{\kappa-i\infty}^{\kappa+i\infty}  \frac{\bar W(p)e^{s p}\prod\limits_{n=0}^{N}(k_n^2-p^2)}{ p\sum\limits_{n=0}^{N}\left\{ C_n \prod\limits_{m\neq n}(k_m^2+p^2)\right\}} dp.
\end{align}
In the limit of $p\to\pm i\infty$, the products in the numerator, as well as the product in the denominator, in the expression under the integral could be modified as
\begin{align}\label{eq:a2}
    \prod\limits_{n=0}^{N}(k_n^2+p^2) &\xrightarrow[p\to \pm i\infty] {}p^{2N}, \\ \nonumber
    p\sum\limits_{n=0}^{N}\left\{ C_n \prod\limits_{m\neq n}(k_m^2+p^2)\right\} &\xrightarrow[p\to \pm i\infty] {} p^{2N-1}\sum\limits_{n=0}^{N} C_n.
\end{align}
Using Equations~\ref{eq:a2}, we find that the expression under the integral in the vicinity of the points $p=\pm i \infty$ simplifies to 
\begin{align}
   \frac{\bar W(p)e^{s p}\prod\limits_{n=0}^{N}(k_n^2+p^2)}{ p\sum\limits_{n=0}^{N}\left\{ C_n \prod\limits_{m\neq n}(k_m^2+p^2)\right\}} \xrightarrow[p\to \pm i\infty]{} \frac{p\bar W(p)e^{s p}}{\sum\limits_{n=0}^{N} C_n}.
\end{align} 
We recall from Eq.~\ref{eq:L1} that 
\begin{align}
\label{eq:L1A}
    \mathscr{L}\left[W'(s) \right]=p\bar W(p)-W(0),
\end{align}
and assuming $W(0)=0$ we arrive at
\begin{align}
\label{eq:rhointA2}
&\rho_z(s)=\frac{W'(s)}{\sum\limits_{n=0}^{N} C_n}+\\ \nonumber &\frac{1}{2 \pi i} \int\limits_{\kappa-i\infty}^{\kappa+i\infty}\left[  \frac{\bar W(p)e^{s p}\prod\limits_{n=0}^{N}(k_n^2+p^2)}{ p\sum\limits_{n=0}^{N}\left\{ C_n \prod\limits_{m\neq n}(k_m^2+p^2)\right\}}-\frac{p\bar W(p)e^{s p}}{\sum\limits_{n=0}^{N} C_n} \right] dp.
\end{align}
\begin{figure}[t]
	\centering
	\includegraphics[width=0.6\linewidth]{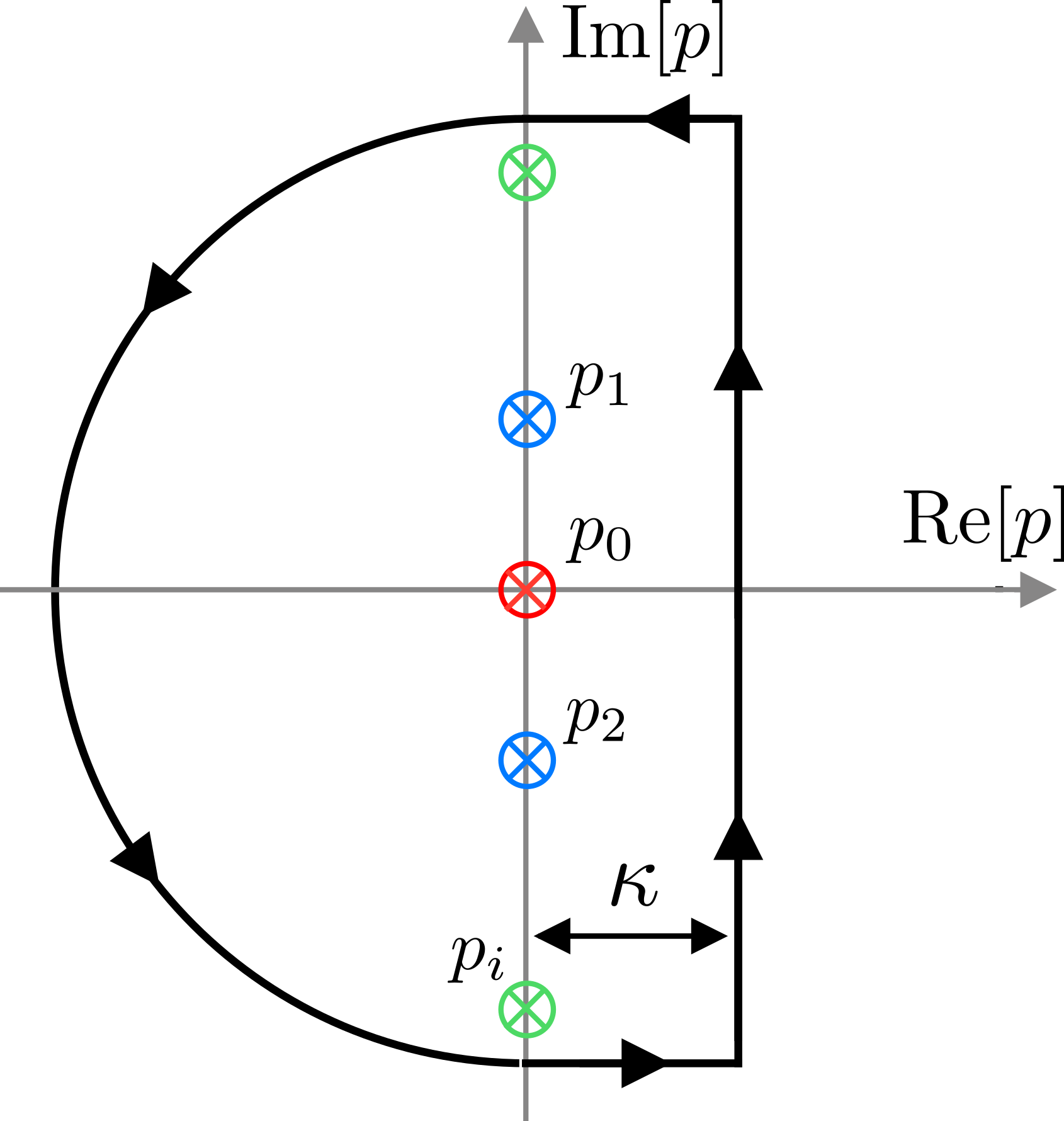}
	\caption{Schematics of the integration contour for the inverse Laplace transformation. }
\label{Fig:LPC}
\end{figure}
The integral in the second term does not have an essential singularity anymore as it is canceled out. 

To evaluate the integral in the Eq.\eqref{eq:rhointA} first we apply convolution theorem for the Laplace transformation  \cite{shabat} in a form
\begin{align}
\mathscr{L}\left[W*K_f\right] = \bar W(p) \bar K_f(p)
\end{align}
with $\bar K_f(p)$ given by
\begin{align}
    \bar K_f(p)=\frac{\prod\limits_{n=0}^{N}(k_n^2+p^2)}{ p\sum\limits_{n=0}^{N}\left\{ C_n \prod\limits_{m\neq n}(k_m^2+p^2)\right\}}-\frac{p}{\sum\limits_{n=0}^{N} C_n}.
\end{align}
This allows one to rewrite Eq.\eqref{eq:rhointA2} as
\begin{align}
\label{eq:rhointA3}
&\rho_z(s)=\frac{W'(s)}{\sum\limits_{n=0}^{N} C_n}+W(s)*K_f(s).
\end{align}
The next step is the evaluation of the integral $K(s)=\mathscr{L}^{-1}\left[\bar K(p) \right]$ 
\begin{align}
\label{eq:Kers}
    &K_f(s)=\\ \nonumber &\frac{1}{2 \pi i} \int\limits_{\kappa-i\infty}^{\kappa+i\infty}\left[  \frac{e^{s p}\prod\limits_{n=0}^{N}(k_n^2+p^2)}{ p\sum\limits_{n=0}^{N}\left\{ C_n \prod\limits_{m\neq n}(k_m^2+p^2)\right\}}-\frac{p e^{s p}}{\sum\limits_{n=0}^{N} C_n} \right] dp.
\end{align}
The integral in Eq.~\eqref{eq:Kers} does not have any singularities in the left half plane (Fig.\ref{Fig:LPC}) except for the poles that are defined be the polynomial in the denominator of the first term. The Corresponding equation for the poles reads
\begin{align}
\label{eq:ple}
    p\sum\limits_{n=0}^{N}\left\{ C_n \prod\limits_{m\neq n}(k_m^2+p^2)\right\}=0.
\end{align}
Therefore the integral in Eq.\eqref{eq:Kers} could be calculated using the residue theorem \cite{shabat,silverman}.  
It is apparent that the second term under the integral in Eq.\eqref{eq:Kers} does not contribute to these residues, thus 
\begin{align}
\label{eq:Kf}
     K_f(s)=\sum\limits_{p_i} \Res \left[ \frac{e^{s p}\prod\limits_{n=0}^{N}(k_n^2+p^2)}{ p\sum\limits_{n=0}^{N}\left\{ C_n \prod\limits_{m\neq n}(k_m^2+p^2)\right\}}\right].
\end{align}
We notice that the pole at $p=0$ could be always evaluated explicitly and reads
\begin{align}
\label{eq:zp}
\Res(p=0) = \frac{\prod\limits_{n=0}^{N}k_n^2}{\sum\limits_{n=0}^{N}\left\{ C_n \prod\limits_{m\neq n}k_m^2\right\}}.
\end{align}
With Eq.\eqref{eq:Kf} and Eq.\eqref{eq:rhointA3} accounting for Eq.\eqref{eq:zp} we finally arrive at
\begin{align}
\label{eq:MMFA}
    \rho_z(s)&=\frac{W'(s)}{\sum\limits_{n=0}^{N} C_n}+\frac{\prod\limits_{n=0}^{N}k_n^2}{ \sum\limits_{n=0}^{N}\left\{ C_n \prod\limits_{m\neq n}k_m^2\right\}}\int\limits_{0}^{s} W(\tilde s) d\tilde s \nonumber \\ &+\int\limits_{0}^{s}K(s-\tilde{s})W(\tilde{s})d\tilde{s}.
\end{align}
Here $K(s)$ denotes the sum of residues that are defined by the roots of Eq.\eqref{eq:ple} except for the trivial root $p=0$: 
\begin{align}
\label{eq:Kff}
     K(s)=\sum\limits_{p_i\neq0} \Res \left[ \frac{e^{s p}\prod\limits_{n=0}^{N}(k_n^2+p^2)}{ p\sum\limits_{n=0}^{N}\left\{ C_n \prod\limits_{m\neq n}(k_m^2+p^2)\right\}}\right].
\end{align}

\bibliographystyle{unsrt}

\end{document}